# Bloch point-mediated skyrmion annihilation in three dimensions

M. T. Birch,[1,2*] D. Cortés-Ortuño,[3] N. D. Khanh,[4,5] S. Seki,[4,6,7] A. Štefančič,[8]† G. Balakrishnan,[8] Y. Tokura,[4,6,9] P. D. Hatton[2]

[1] Max Planck Institute for Intelligent Systems, 70569 Stuttgart, Germany.

[2] Department of Physics, Durham University, Durham DH1 3LE, United Kingdom.

[3] Department of Earth Sciences, Utrecht University, 3584 CB Utrecht, The Netherlands.

[4] RIKEN Center for Emergent Matter Science (CEMS), Wako, 351-0198, Japan.

[5] Institute for Solid State Physics, The University of Tokyo, Chiba 277-8581, Japan.

[6] Department of Applied Physics, The University of Tokyo, Bunkyo, Tokyo 113-8656, Japan.

[7] PRESTO, Japan Science and Technology Agency (JST), Kawaguchi 332-0012, Japan.

[8] Department of Physics, University of Warwick, Coventry, CV4 7AL, United Kingdom.

[9] Tokyo College, The University of Tokyo, Bunkyo, Tokyo 113-8656, Japan.

† Current address: Electrochemistry Laboratory, Paul Scherrer Institut, CH-5232 Villigen PSI, Switzerland.

The creation and annihilation of magnetic skyrmions are mediated by three dimensional topological defects known as Bloch points. Investigation of such dynamical processes is important both for understanding the emergence of exotic topological spin textures, and for future engineering of skyrmions in technological applications. However, while the annihilation of skyrmions has been extensively investigated in two dimensions, in three dimensions the phase transitions are considerably more complex. We report field-dependent experimental measurements of metastable skyrmion lifetimes in an archetypal chiral magnet, revealing two distinct regimes. Comparison to supporting three-dimensional geodesic nudged elastic band simulations demonstrates that these correspond to skyrmion annihilation into either the helical or conical states, each exhibiting a different transition mechanism. The results highlight that the lowest energy magnetic configuration of the system plays a crucial role when considering the emergence of topological spin structures via defect-mediated dynamics, and their stability in future devices.

## Introduction

Topological defects have been shown to mediate phase transitions by breaking local symmetries. For example, Kosterlitz and Thouless suggested that the proliferation and propagation of dislocations and disclinations, respectively breaking translational and rotational symmetry, are responsible for the melting of a 2D crystalline solid into a liquid (*1*). Such discontinuities are excited states, and can be created either in opposing pairs within the system, or singly at the system's boundaries (*2*). Topological defects have been observed in a range of physical systems, such as in the nematic phase of liquid crystals (*3*), as fluxons in type-II superconductors (*4*), or as magnetic vortices in the two dimensional X-Y spin model (*1*).

Topological phases have been the focus of recent research due to their unique behaviors and properties. In particular, magnetic skyrmions, stabilized by the Dyzloshinskii-Moriya interaction (DMI), have been investigated due to their potential applications in future spintronic devices (*5-7*). Similar to the melting of solids by dislocations and disclinations, magnetic skyrmions are annihilated by the proliferation and subsequent propagation of magnetic Bloch points (*8*). Such three dimensional topological defects have been called hedgehog defects, or emergent magnetic monopoles (*9*) due to their singular nature. Finding methods of controlling this dynamic behavior is crucial for realizing reading and writing protocols in proposed skyrmion devices (*10-12*).

Due to the focus on thin-film applications, previous investigations into the stability and annihilation mechanisms of skyrmions have largely focused on isolated skyrmions in two dimensional systems (*13-15*). In these confined systems, the energy barrier required to annihilate a skyrmion at the boundary of the system is typically lower than that required for its destruction by direct collapse (*16, 17*). Further calculations have demonstrated that entropic considerations, which concern the number of available paths across the annihilation energy barrier, play a more prominent role in determining the lifetime of skyrmions in two dimensional systems than the energy barrier height itself (*18-21*). This phenomenon has been demonstrated experimentally (*18, 22*).

However, while they are commonly portrayed as two dimensional objects, in three dimensions, magnetic skyrmions exist as extended, tube-like objects (*23, 24*). In a bulk helimagnetic skyrmion system, such a skyrmion tube (SkT) can exist within either a helical or conical state, as shown by the visualizations in Fig. 1A and B. In each configuration, the SkT is annihilated by a different Bloch point-mediated mechanism, as depicted in Fig. 1C. On the left, the SkT connects with the edge of the nearby helical domain, forming a pair of Bloch points ($H^+$ and $H^-$), whose subsequent motion joins the SkT to the neighboring helical domain (*8*). On the other hand, on the right side, the SkT breaks in two, again forming a pair of Bloch points ($C^+$ and $C^-$), which unwind the SkT into the conical state (*25*). Cross sections of the system are shown in Fig. 1D, including insets showing the spin configuration surrounding each Bloch point.

While there have been previous computational studies into the topological phase transitions governing three dimensional skyrmion annihilation (*26-29*), a combined experimental and computational investigation into the decay dynamics in a bulk skyrmion system has yet to be performed. Beyond being vital for technological implementations, such three dimensional Bloch point-mediated mechanisms likely hold the key to understanding the stabilization of more exotic topological phases, for example: the emergence of the low temperature skyrmion phase from the tilted conical state in $Cu_2OSeO_3$ (*30, 31*), the transformation of hexagonal skyrmion lattices to meron-antimeron lattices in $Co_8Zn_9Mn_3$ (*32*), or future experimental realization of magnetic hopfion states (*33*).

## Results

### Metastable Skyrmion Phase Diagram

In bulk chiral crystals, skyrmions typically exist in a small range of temperature and applied magnetic field close to the Curie temperature, $T_C$, of the material (*34*). Previous studies have shown that a metastable skyrmion state can be realized outside this equilibrium region by cooling the sample under an applied magnetic field (*35-37*). Being metastable, this state has a finite, temperature dependent lifetime (*38*), which can be studied with time-resolved measurements (*39*). It has been noted that disorder and defects in the underlying crystal lattice increase the lifetime of the metastable skyrmion state (*35*). Therefore, in order to stabilize metastable skyrmions over the widest possible range of temperature and applied magnetic fields, we chose to investigate a Zn-substituted $Cu_2OSeO_3$ sample, which exhibits such an enhanced lifetime (*40*).

Figure 2A shows the magnetic phase diagram of the chosen $(Cu_{1-x}Zn_x)_2OSeO_3$ ($x=0.02$) single crystal, with a $T_C = 57.5$ K (see Supplementary Fig. S1). Being multiferroic, the phase diagram was determined by measurements of the electric polarization, $P$, along the [001] crystal axis, as a function of the magnetic field applied along the [110] axis (*41*) (see Supplementary Fig. S2). At low fields, the ground state consists of helical domains, with their spin modulations oriented along the ⟨001⟩ axes due to the cubic anisotropy (*42*). With increasing applied field, the conical state is realized, with a propagation vector aligned with the magnetic field direction. The phase diagram illustrates the stabilization of the equilibrium skyrmion lattice (SkL) state over a short range temperature and applied field at ~56 K, and the comparably larger extent of the metastable SkL state, realized by field cooling at 20 mT.

Corresponding AC susceptibility measurements are displayed in Fig. 2B and 2C, performed after zero field cooling (ZFC), high field cooling (HFC) and field cooling (FC) the sample to the target temperature (see Methods). In Fig. 2B, the decrease in the real component of the AC susceptibility, $\chi'$, around 20 mT is characteristic of the formation of the SkL state (*43*), as highlighted in yellow. Upon FC at 20 mT to 48 K, a similar characteristic decrease in $\chi'$ is indicative of the presence of metastable skyrmions, as shown in Fig. 2C.

Examination of the magnetic phase diagram reveals that the metastable SkL state overlays both the conical and the helical ground states, as indicated by the yellow hatched and dotted sections on the phase diagram. Therefore, one might expect that the metastable SkL state will decay into either the conical (SkL→C) or helical (SkL→H) states depending on the field applied to the sample. This affords us the opportunity to study the dynamics of both skyrmion annihilation mechanisms proposed in Fig. 1.

### Metastable Skyrmion Lifetime

Measurements of the AC susceptibility as a function of time allow the annihilation rate of the metastable skyrmion state to be measured. Specifically, we can expect that the difference in the value of $\chi'$ between the FC and ZFC/HFC measurements is proportional to the population of metastable skyrmions present in the sample (*40, 44*). Thus, when the metastable skyrmions annihilate, the value of $\chi'$ will relax over time, as indicated by the vertical arrow in Fig. 2C. We carried out such time-resolved skyrmion lifetime measurements as a function of the applied magnetic field. A selection of the resulting data is displayed in Fig. 3A (full dataset shown in Supplementary Fig. S3). Here, the AC susceptibility has been normalized, $\chi'_N = (\chi'_f - \chi'(t))/(\chi'_f - \chi'_0)$, where $\chi'_0$ is the initial value of $\chi'$ at $t = 0$, and $\chi'_f$ is the value the system is tending to as $t \rightarrow \infty$.

The lifetime of the metastable skyrmion state at each field, $\tau$, was extracted by fitting the data to a stretched exponential decay function (*45*),

$$\chi'_N(t) = -\exp\left[-\left(\frac{t}{\tau}\right)^\beta\right]. \tag{1}$$

A stretching parameter of $\beta \neq 1$ indicates that the system displays a range of lifetimes. Such a distribution of lifetimes may be due to inhomogeneous magnetic fields within irregularly shaped samples (*46, 47*). Nevertheless, for our measurements, the inclusion of $\beta$ is necessary to fit the data, and varies between 0.3 and 0.6 for different applied fields.

The resulting metastable skyrmion lifetimes are plotted as a function of magnetic field in Fig. 3B. The lifetime is at a maximum at 24 mT, and decreases at magnetic fields above and below this point. The two distinct exponential trends point towards the observation of the two anticipated regimes of the SkL→H and SkL→C annihilation mechanisms, and we tentatively label the corresponding regions of the figure accordingly.

A modified Arrhenius' law can be employed to describe the temperature dependence of the skyrmion lifetimes, allowing the energy barrier, $E_B$, governing the decay mechanism to be determined,

$$\tau(T) = \tau_0 \exp\left[\frac{-E_B}{k_B T}\right] = \tau_0 \exp\left[a\frac{(T_s - T)}{T}\right]. \tag{2}$$

Here, the temperature dependence of the energy barrier is assumed to be approximately linear, with $a$ as the proportional constant, following the relationship $E_B/k_B = a(T - T_s)$ (*39*). The lifetime can be expected to reach a minimum at the lowest temperature extent of the equilibrium skyrmion phase, $T_s$, which we determined to be approximately 4 K below $T_C$ (see Supplementary Figure S1).

Considering equation (2), the two distinct exponential trends in Fig. 3B suggest both a negative or positive linear dependence of $a$ with the applied magnetic field. However, one must also consider the contribution of the $\tau_0$ term. This prefactor is typically thought of as a characteristic attempt frequency with which the system attempts to overcome the energy barrier. The term also includes the aforementioned entropic correction which has been shown to play a crucial role in the stability of magnetic skyrmions in thin lamellae (*22*).

In order to separate the contributions of $a$ and $\tau_0$, we performed further, temperature dependent lifetime measurements at five applied magnetic fields between 16 and 34 mT (full data shown in Supplementary Figure S3). The resulting extracted lifetimes are plotted as a function of temperature in Fig. 3C. By fitting these lifetimes with the modified

Arrhenius law in equation (2), the values of $a$ (the gradient) and $\tau_0$ (the value of $\tau_0$ at $T_s$) were determined, and are plotted in Fig. 3D.

Immediately, one can see that $a$ appears to decrease either side of 24 mT, as was indicated by the initial lifetime measurements in Fig. 3B. This supports the interpretation that the energy barrier appears to vary approximately linearly, with either a positive or negative gradient, in two regimes which correspond to the SkL→H and SkL→C transitions respectively. However, the contribution from the $\tau_0$ term appears to decrease exponentially across the measured field range. This relationship is significantly different to the one determined by Wild *et al.* in their thin lamella sample (*22*). Following the phenomenological Meyer-Neldel compensation rule (*21*), they demonstrated that $\tau_0$ varied exponentially with the measured energy barrier - concluding that the entropy therefore varies linearly with the energy barrier. However, when plotting our measured $a$ and $\tau_0$ values in the inset of Fig. 6D, it is clear that such a dependence is not recovered for our bulk sample, and entropic considerations therefore appear to play a less prominent role.

The discrepancy can be explained by considering that in measurements on three dimensional systems, the annihilation of skyrmions consists of two processes: the nucleation of Bloch point pairs, and the subsequent motion of these Bloch points along the length of the skyrmion tube. In such bulk systems, $\tau_0$ appears to be correlated with increased disorder and defects within the underlying crystal lattice (*44*). Therefore, the decrease of $\tau_0$ with increasing magnetic field in our results indicates that the thermally-activated depinning of the Bloch-point from structural defects becomes more energetically favorable at higher applied magnetic fields. Any entropic contributions may well be obscured by these pinning effects.

**Energy Barrier Simulations**

To support the interpretation of the experimental measurements, we modelled a chiral magnet by means of a classical spin model with periodic boundaries in the *xy* plane and open boundaries in the *z* direction (see Methods). To begin, suitable magnetic configurations for the helical and conical states were prepared and their energy was numerically minimized over the field range. This was repeated for single SkTs embedded within helical and conical states, resulting in the configurations depicted in Fig. 1A and B (see Supplementary Fig. S4, S5).

To investigate the transitions of isolated SkTs into helical and conical states, the geodesic nudged elastic band method (GNEBM) was employed (*13, 48*). The GNEBM finds minimum energy paths (MEPs) through configuration space between two equilibrium states by minimizing the energy of a chain of copies of the system known as images. This allows the determination of the energy barrier for the transition (see Methods).

We first applied this algorithm to determine the MEP for a SkT annihilating within a helical state (SkL→H) for applied magnetic fields between $b_z = 0.04$ and $0.16$ (in units of exchange, see Methods), as shown in Fig. 4 (full results shown in Supplementary Fig. S6, Supplementary Movie 1). Since previous measurements suggest that the skyrmion state prefers to annihilate into helical domains with magnetic modulation in the same plane as the skyrmion lattice state (*8, 49*), we chose to utilize helical domains aligned along the *x* axis accordingly. We found it was necessary to apply a cubic anisotropy of $k_c = 0.05$ in order to stabilize the helical domains against the conical state over the investigated field range.

As shown by the selected visualizations of the simulated images at the bottom of Fig. 4A, the SkT decays by forming a pair of Bloch points, whose subsequent motion connects the SkT and the helical domain. The top panel plots the energy of the system along the determined MEP, parameterized by the reaction coordinate, $X$, through configuration space, where each data point is associated with an image of the system. The trajectories of the Bloch points along the $z$-axis are plotted in the middle panel, with the vertical dashed lines indicating images where Bloch points are created or destroyed. By comparison of these panels, one can see that the two energy barriers along the MEP are each associated with the formation of a pair of Bloch points.

We repeated the GNEBM simulations at a range of applied magnetic fields. At some applied fields, both pairs of Bloch points form and annihilate simultaneously, as shown by the simulation performed at $b_z = 0.10$ in Fig. 4B. Furthermore, at $b_z = 0.14$ in Fig. 4C, we found a simultaneous formation of four pairs of Bloch points. At $b_z = 0.14$ and 0.16, the final state was found to be higher in energy compared to the SkT in helical state, suggesting that, for the chosen parameters, the SkT within the helical state was the lowest energy configuration over a small range of field. Due to the complexity of the transition, there are likely many similar paths through configuration space, all with energy close to the MEP, and therefore we speculate that the formation of different numbers of Bloch points may not be correlated to the field applied to the system.

There has been a previous work utilizing Ginzburg-Landau analysis to investigate energy barriers for a SkT decaying within a helical state. However, the decay mechanism the authors found was more akin to the breaking of the SkT within a conical state (*26*), rather than the joining of the SkT to the helical domains as seen in this work, and other experimental works (*8, 22*).

We performed similar GNEBM simulations for the SkT in a conical state (SkL→C) at applied fields between $b_z = 0.36$ and 0.50, as shown in Fig. 5 (full results shown in Supplementary Fig. S7, Supplementary Movie 2). With the same cubic anisotropy utilized for the helical simulations, we found that the conical state was only the lowest energy state over a short range of applied field (see Supplementary Fig. S8, Supplementary Movie 3). Therefore, in order to simulate the SkL→C transition over a wide field range, we performed this set of simulations with no cubic anisotropy present.

Fig. 5A shows the results of the SkL→C simulation under an applied field of $b_z = 0.36$. In comparison to the SkL→H simulations, the energy landscape is considerably more complex, with many more metastable states possible along the MEP. At the bottom of panel A, images of the system along the MEP show the breaking of the SkT into three sections by the formation of two pairs of Bloch points. The sections consist of two chiral bobbers at the open boundary surfaces of the system (*50*), and a short SkT capped by two Bloch points, known as a toron (*51, 52*). Following along the MEP, the formation of each Bloch point pair is once again associated with an energy barrier, as shown in the top panel. In addition, at this lower field there is an energy barrier associated with the collapse of the toron (image 8), and the two chiral bobbers (images 10 and 11), suggesting they are metastable states. We confirmed that these states are indeed stable local minima by energy minimization.

At $b_z = 0.48$ and above, the applied field is sufficient to transform the conical state into the uniformly magnetized state. From this point, only one pair of Bloch points is formed during the SkT annihilation, as shown in Fig. 5C, which resembles previous computational works (*27, 28*). Furthermore, there is no longer an energy barrier associated with the destruction of the torons and chiral bobbers, suggesting they are not metastable configurations at higher applied fields.

To allow comparison of these simulations with the experimental results, we calculated the energy barrier required for the formation of the Bloch point pairs in each simulation, as indicated by the arrow in Fig. 4B. In cases where two, or four, pairs of Bloch points were formed simultaneously, we divided the energy by the number of created pairs. Energy barriers associated to the formation of Bloch point pairs for the SkL→H and SkL→C transitions are shown as a function of applied field in Fig. 6A and B respectively. In addition, energy barriers for the formation of Bloch point pairs for SkT tubes annihilating within the uniformly magnetized state (SkL→UM) with cubic anisotropy of $k_c = 0.05$ are shown in Fig. 6C (see Supplementary Figs S8). Finally, energy barriers associated with the annihilation of toron and chiral bobber states within the conical and uniformly magnetized simulations are shown in Fig. 6D.

**Discussion**

Considering Fig. 6A and B, one can see that the energy barrier associated with Bloch point formation increases as a function of the applied field for the SkT within the helical state, whereas it decreases as a function of field for the SkT within the conical state. This qualitatively agrees with the conclusions drawn from the experimental data in Fig. 3, where we anticipated the energy barriers varying linearly with the applied field, supporting our labelling of the two linear regimes in the lifetime data to the two distinct SkT annihilation mechanisms.

Our results of SkT annihilation to the conical state can be compared to previous results of isolated skyrmions in two dimensional systems, which typically show two transition paths: either the skyrmion is annihilated at the boundary of the system, or it shrinks and collapses by flipping the central spin (*16*). The latter resembles our results in Fig. 5, where the SkT narrows prior to the nucleation of the Bloch point pair, at the point where the spin in the center of the skyrmion flips to align with the applied field. We also investigated the possibility of surface Bloch point nucleation (see Supplementary Fig. S10, Supplementary Movie 4), but when applying the GNEBM, the algorithm appeared to prefer forming Bloch point pairs within the interior of the system over the simulated field range.

To summarize, our experimental measurements of field-dependent metastable skyrmion lifetimes indicated two distinct regimes, where skyrmion annihilation is dominated by decay into either the helical or conical states. This conclusion is supported by GNEBM simulations, where the determined energy barriers associated with Bloch point-mediated skyrmion annihilation to the helical and conical states revealed the same field dependence. In contrast to previous studies of two-dimensional systems, in our bulk sample the effects of defect pinning appear to dominate over entropic contributions. The results highlight that, for technological applications, consideration of the competing magnetic ground state and the dimensionality of the system will be crucial when engineering skyrmion stability for reading and writing processes. Further utilization of the GNEBM, and similar computational methods, will no doubt prove to be a powerful tool in the understanding of topological defect-mediated phase transitions, particularly when considering the stabilization of future complex three dimensional spin textures.

## Materials and Methods

### Sample Growth and Preparation

A single crystal of $(Cu_{1-x}Zn_x)_2OSeO_3$ was grown from polycrystalline powders using the chemical vapor transport method (*53*). The composition of the resulting crystal was determined by inductively-coupled plasma mass spectroscopy, and energy-dispersive X-ray spectroscopy, yielding $x = 0.02$ (2% Zn substitution), as reported in previous work (*53*). For electric polarization measurements, the sample was aligned and cut into a plate shape, with two large parallel ⟨111⟩ faces.

### AC Susceptibility Magnetometry

AC susceptibility measurements were performed with a MPMS3 Quantum Design magnetometer, fitted with the AC option. The sample was attached to a quartz rod using GE varnish and mounted in the instrument with the [110] crystal direction parallel to the applied magnetic field. AC susceptibility measurements were performed with an oscillating magnetic field amplitude of 0.1 mT and a frequency of 10 Hz. All cooling procedures were performed at a rate of 40 K/min. For the ZFC, FC and HFC measurements, the sample was cooled from 60 K to the target temperature under an applied field of 0, 20 or 200 mT respectively, and measurements subsequently performed as a function of increasing or decreasing magnetic field. For the lifetime measurements, the sample was initialized following the FC process, and the AC susceptibility was then measured as a function of time under stable conditions for three or more hours.

### Electric polarization Measurements

For the evaluation of electric polarization, *P*, silver paste was painted on a pair of (001) surfaces of the $(Cu_{1-x}Zn_x)_2OSeO_3$ crystal. With the field applied along the [110] crystal axis, We measured the pyroelectric current during a constant rate of magnetic field sweep using an electrometer (Keithley 6517B), and integrated over time to deduce *P* (*41*).

### Computational Modelling

The system was modelled using a cubic lattice of N = 30×30×30 classical spins with periodic boundaries in the *xy* plane. The system follows a Heisenberg-like Hamiltonian where energy constants of the different interactions are made dimensionless by normalizing by the exchange constant,

$$\frac{H}{J} = -\sum_{<i,j>}^{N} \mathbf{s}_i \cdot \mathbf{s}_j + d \sum_{<i,j>}^{N} \mathbf{r}_{ij} \cdot [\mathbf{s}_i \times \mathbf{s}_j] - k_c \sum_i^N (s_{x,i}^4 + s_{y,i}^4 + s_{z,i}^4) - \sum_i^N b_z s_{z,i}.$$

Here, $\mathbf{s}_i$ are unit vectors denoting spin orientations, $d = D/J > 0$ is the DMI constant for a cubic helimagnet, $k_c = K_c/J > 0$ is the cubic anisotropy constant (easy ⟨100⟩ axes) and $b_z = (\mu_s/J)B_z$ is the *z*-component of the applied field which is specified in units of magnetic moment per exchange constant $\mu_s/J$. The angled brackets in the first two sums denote counting pairs of spins only once.

The magnitude of $d \approx 0.727$ was set such that the periodicity of a spin spiral at zero field and in the absence of any anisotropy is 10 lattice sites (*51, 54, 55*). The value of the saturation field in absence of anisotropy is defined as $B_D = D^2/(\mu_s J) = d^2 \approx 0.53$ (*55*). To stabilize helical configurations, a cubic anisotropy constant of $k_c = 0.05$ was chosen, while for the conical simulations, $k_c = 0.00$ was specified. Equilibrium states were obtained by initializing the system with a suitable configuration and minimizing the energy using an over-damped Landau-Lifshitz-Gilbert (LLG) equation (see Supplementary Fig. S4, S5).

The GNEBM finds minimum energy paths (MEPs) through configuration space between two magnetic states (*13*). Using a suitable initial state, a series of images along the transition path are calculated. During the algorithm, the images follow gradients of energy and are kept distanced in configuration space until reaching a MEP between the initial and final states. Saddle points along the MEP determine the energy barriers of the transitions. In this study we use a rotation formula for the spin orientations to initiate the algorithm (*13*).

All simulations and GNEBM calculations were obtained using Fidimag (*16, 48*). Further details about the simulations and the applied convergence criteria to stop the energy minimization both for the LLG equation and GNEBM, are discussed in the Supplementary Material.

**Acknowledgments**

**General:** The authors thank P. Steadman, R. Fan and M. Sussmuth at Diamond Light Source for help with the magnetometry measurements. Appreciation is given to K. Matsuura and K. Vilmos for assistance with the electric polarization measurements. **Funding:** The work was funded by the UK Skyrmion Project EPSRC Programme Grant (EP/N032128/1). M.T.B. acknowledges support from the Max Planck Society. **Author contributions:** M.T.B., D.C.-O. and P.D.H. conceived the project. A.S. and G.B. fabricated the single crystal sample. N.D.K., S.S. and G.B. performed the electric polarization measurements. M.T.B carried out the magnetometry experiments. D.C-O. performed the computational simulations. M.T.B. and D.C-O. wrote the manuscript with input from all authors. All authors discussed the results. **Competing interests:** The authors declare no competing interests. **Data and materials availability:** All experimental data, and all the code to reproduce the simulations can be found in an online repository (*available soon, 56*).

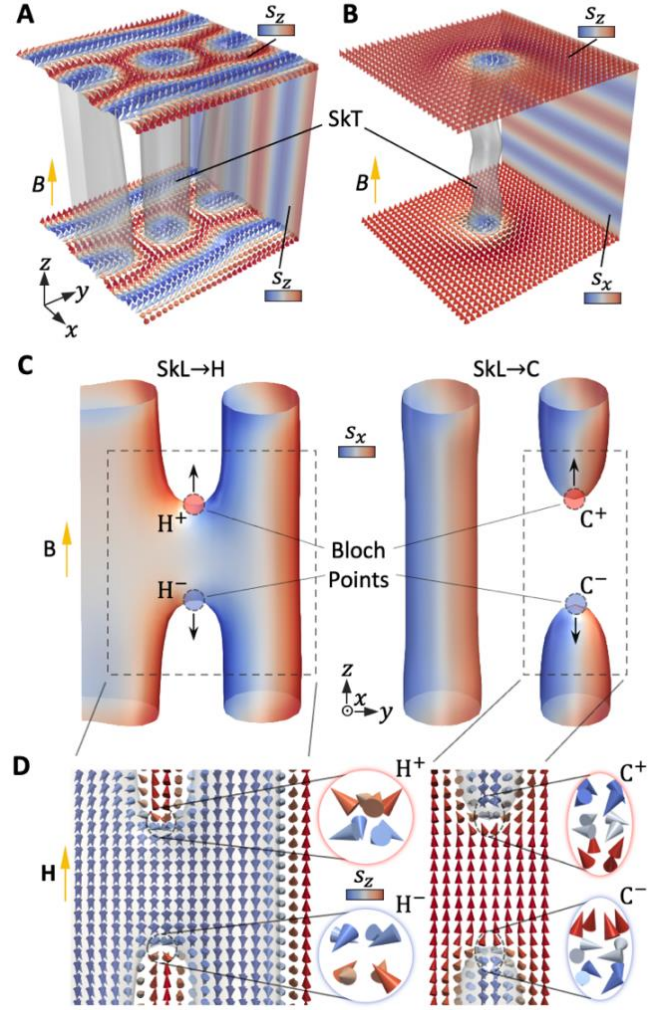

**Fig. 1: Bloch-point-mediated skyrmion annihilation mechanisms**. (**A**, **B**) Three dimensional visualizations of simulations display a skyrmion tube (SkT) embedded in the helical and conical states respectively. The grey contours highlight regions where the magnetization along the $z$ axis, $s_z = 0$. The top and bottom layers of spins have been colored according to their $s_z$ component. The back right surface of each simulation has been colored according to the local $s_z$ or $s_x$ components respectively, highlighting the orientation of the surrounding helical and conical structures. (**C**) Visualizations of the skyrmion to helical (SkL→H), and skyrmion to conical (SkL→C) annihilation mechanisms, where the contour surfaces highlight regions where $s_z = 0$, and the coloration indicates the local $s_x$ component. A pair of Bloch points is nucleated where either the SkT connects to the local helical structure (left, $H^+$ or $H^-$) or breaks in two to form the conical state (right, $C^+$ or $C^-$). (**D**) Cross sections of the spin texture around the Bloch point structures shown in (**C**). Insets display the local spin arrangement around each Bloch point. The colors of the spins indicate the $s_z$ component.

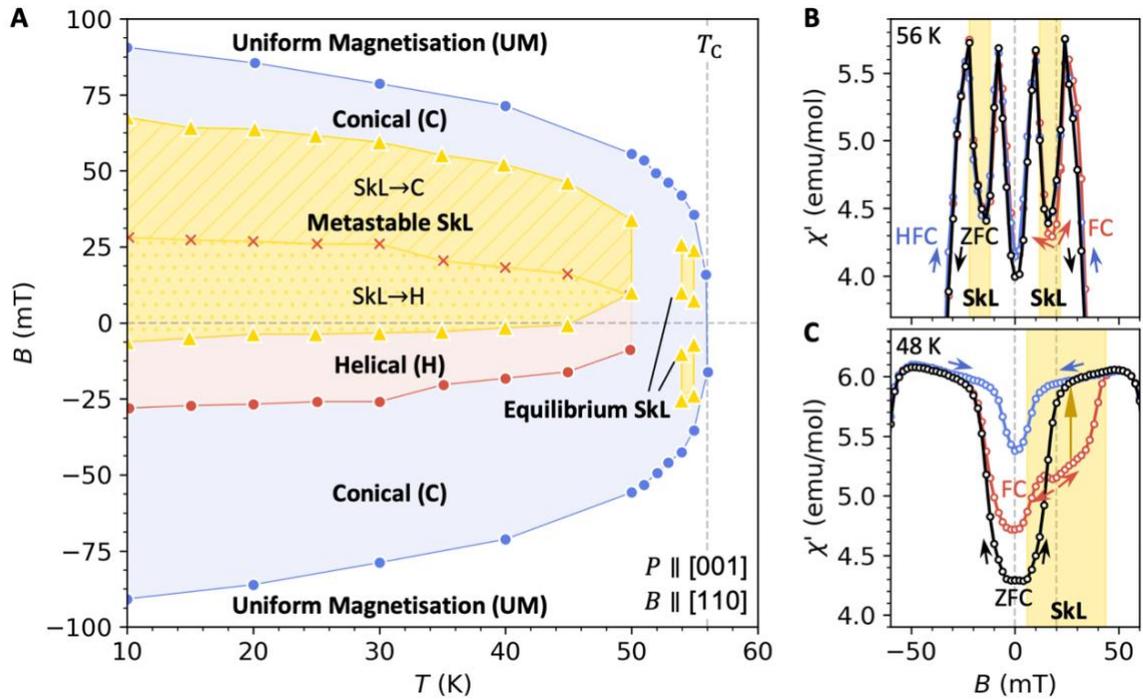

**Fig. 2: Metastable skyrmion phase diagram.** (**A**) The magnetic phase diagram of the $(Cu_{1-x}Zn_x)_2OSeO_3$ sample, as determined by measurements of the electric polarization $P$ along the [001] crystallographic axis, when the magnetic field is applied along the [110] axis. The uniformly magnetized (UM, white), conical (C, blue), helical (H, red) and equilibrium skyrmion lattice (SkL, yellow) phases are labelled. The metastable SkL phase is divided into two regions, where it overlays the equilibrium conical (yellow hatched) and helical (yellow dotted) phases. (**B**, **C**) The real component of the AC susceptibility data, $\chi'$, measured after ZFC (black), HFC (blue) and FC (blue). The decrease in $\chi'$ characteristic of the formation of skyrmions is highlighted by the yellow fill.

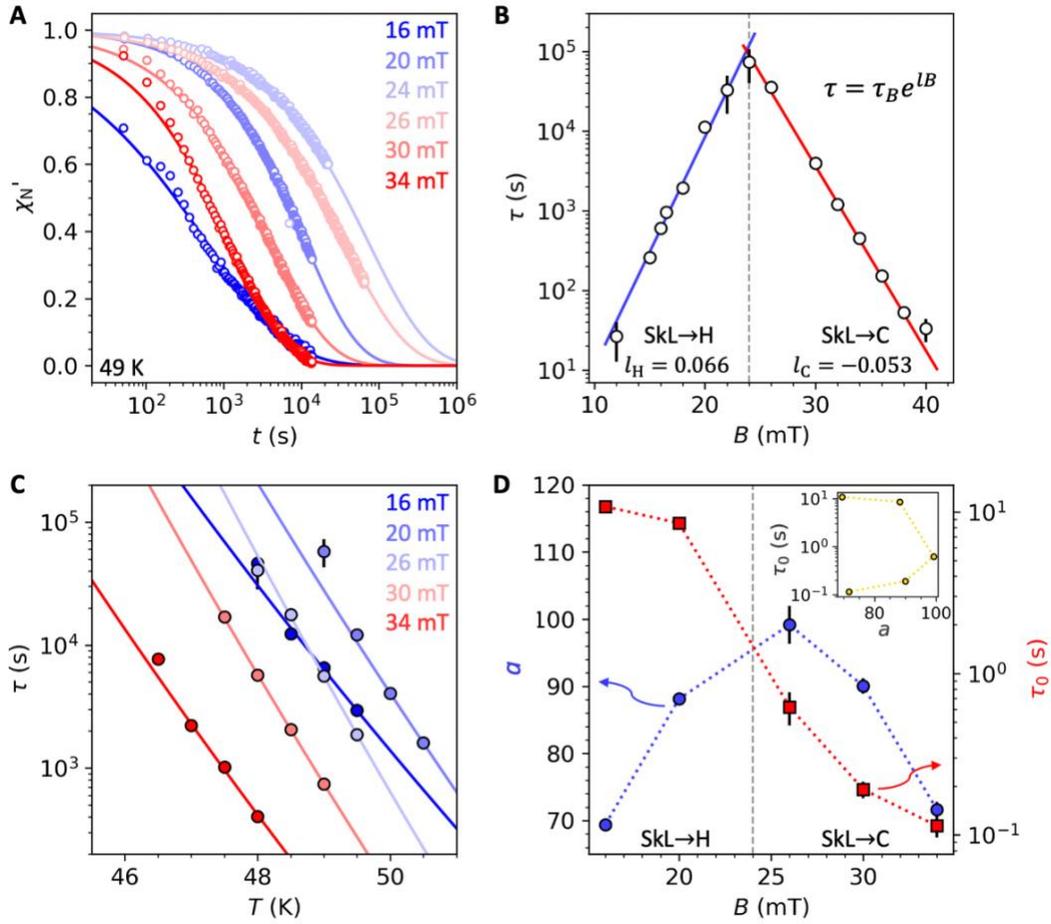

**Fig. 3: Measuring field-dependent skyrmion lifetimes.** (**A**) The normalized real component of the AC susceptibility, $\chi'_N$, measured as a function of time at a range of applied magnetic fields after FC the sample at 20 mT from 65 to 49 K. The data is fitted with the stretched exponential decay function, and the corresponding magnetic fields are labelled. (**B**) The fitted lifetimes are plotted as a function of the applied magnetic field on a logarithmic axis. The two distinct exponential trends are fitted to $\tau = \tau_B e^{lB}$ where $\tau_B$ is the lifetime at 0 mT, and $l$ is the linear constant in the exponent, to determine how the lifetime varies with the applied field. The extracted parameters for $l$ are 0.066 and -0.053 mT$^{-1}$ for the SkL→H and SkL→C ranges respectively. (**C**) Further lifetimes measured at different fields are plotted as function of temperature. The datasets are fitted with a modified Arrhenius law to extract $a$ and $\tau_0$. (**D**) The value of $a$ and $\tau_0$ are plotted as a function of the applied field. In the inset, the determined $\tau_0$ and $a$ parameters are plotted against one another to test the Meyer-Neldel compensation rule, where one expects $\tau_0$ to vary exponentially with the energy barrier $a$.

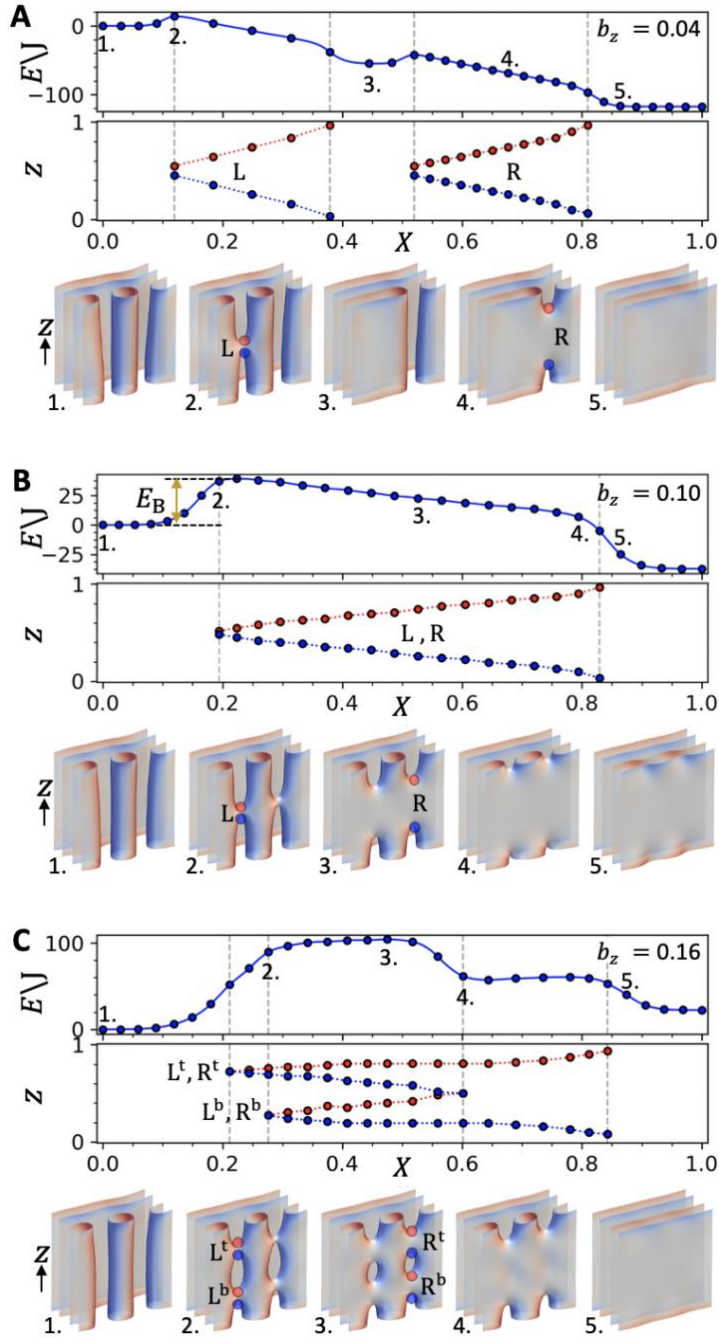

**Fig. 4: Skyrmion to helical annihilation energy barrier simulations.** (**A-C**) The results of GNEBM simulations for the SkL→H decay mechanism are shown at a range of applied magnetic fields. The top panels display the energy dependence of the magnetic state as a function of the reaction coordinate $X$ along the MEP. The middle panels plot the $z$-axis trajectory of the left (L) and right (R) pairs of Bloch points along as a function of $X$ as the SkT is annihilated. Vertical dashed lines highlight images where Bloch points are created or destroyed. The bottom panels exhibit selected three dimensional visualizations of the simulation at specific points along the MEP, as indicated by the numeric labels in the top panels. The surfaces display $s_z = 0$ contours, and have been colored according to the local value of $s_x$.

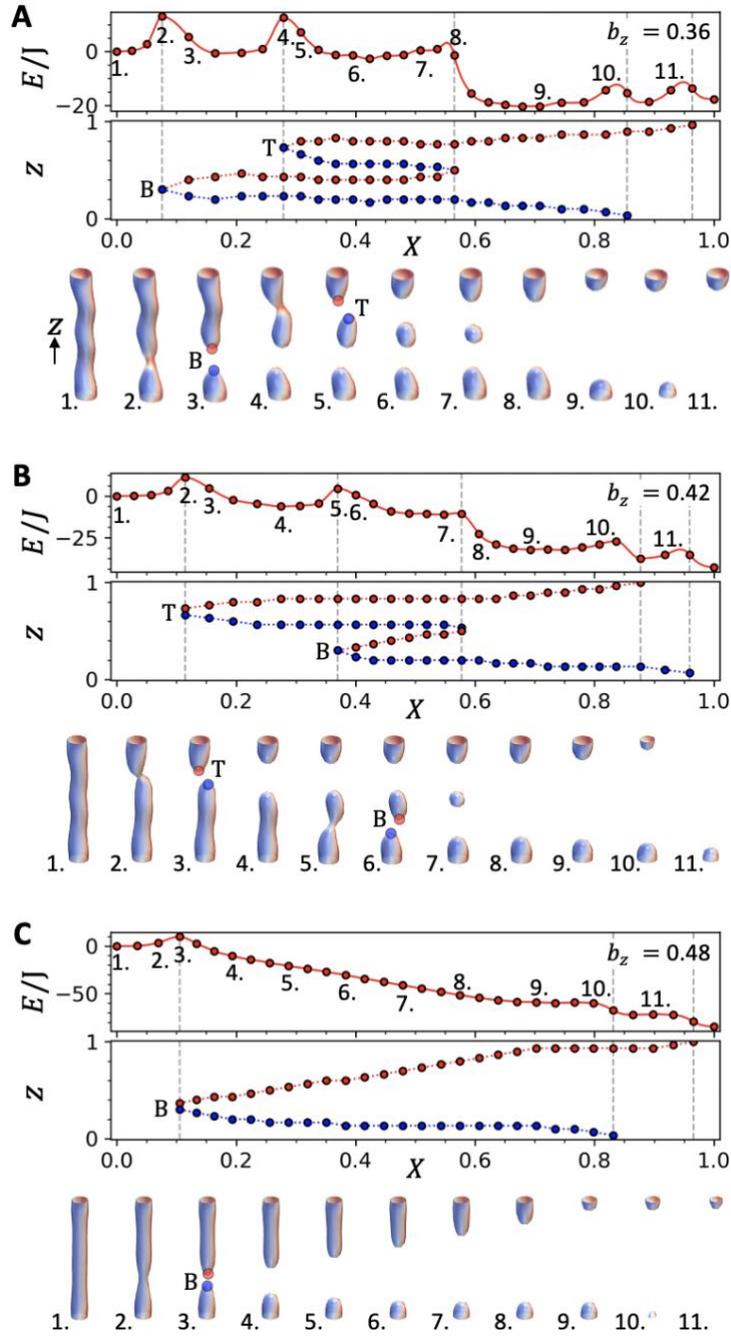

**Fig. 5: Skyrmion to conical annihilation energy barrier simulations.** (A-C) The results of GNEBM simulations for the SkL→C decay mechanism are shown at a range of applied magnetic fields. The top panels display the energy dependence of the magnetic state as a function of the reaction coordinate $X$ along the MEP. The middle panels plot the $z$-axis trajectory of the top (T) and bottom (B) pairs of Bloch points along as a function of $X$ as the SkT is annihilated. Vertical dashed lines highlight images where Bloch points are created or destroyed. The bottom panels exhibit selected three dimensional visualizations of the simulation at specific points along the MEP, as indicated by the numeric labels in the top panels. The surfaces display $s_z = 0$ contours, and have been colored according to the local value of $s_x$.

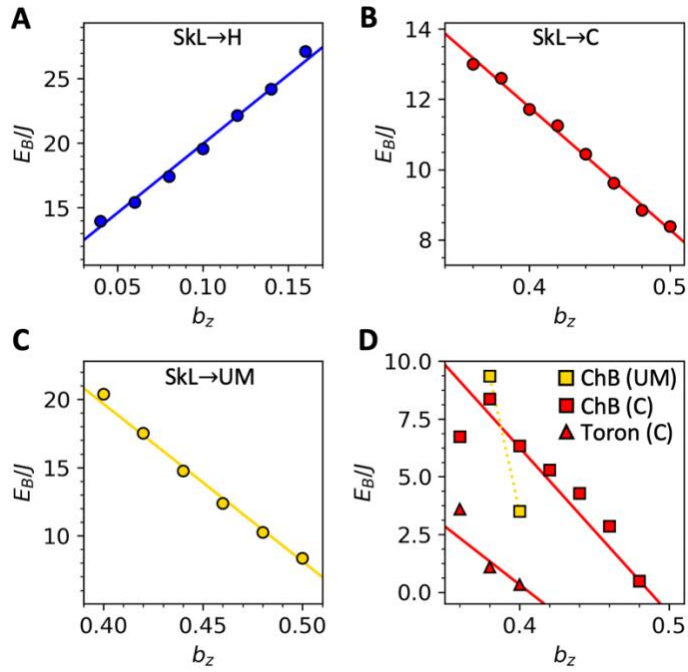

**Fig 6: Field dependence of the annihilation energy barriers.** (**A**-**C**) The field dependence of the simulated energy barrier which must be overcome to form one pair of Bloch points is plotted as a function of the applied field for the SkL→H (blue), SkL→C (red) and SkL→UM (yellow) simulations. (**D**) The simulated energy barrier for the annihilation of a chiral bobber (ChB, squares) or a toron (triangles) is plotted as a function of the applied magnetic field for the SkL→C (red) and SkL→UM (yellow) decay mechanisms. Each dataset has been fitted with a linear trend.

## Supplementary Materials

### Sample Characterization

The $T_C$ of the $(Cu_{1-x}Zn_x)_2OSeO_3$ single crystal sample was determined by measurements of the magnetization as a function of temperature at 20 mT, as shown in Fig. S1A. This data was numerically differentiated, and again plotted as a function of temperature, as shown in Fig. S1B. The peak in $dM/dT$ reveals that $T_C = 57.5$ K. Basic magnetic phase diagrams were determined by measuring AC susceptibility as a function of the applied magnetic field after zero field-cooling (ZFC) and field-cooling at 20 mT (FC). The real component, $\chi'$, is plotted as a colormap in Fig. S1C and D, revealing the reduction in $\chi'$ signal associated with the helical, SkL and metastable SkL states. The corresponding imaginary component, $\chi''$, colormaps are plotted in Fig. S1E and F, revealing the $\chi''$ signal peaks associated with dynamics of the helical to conical and conical to SkL phase transitions.

### Electric Polarization Measurements

Measurements of the electric polarization, $P$, were utilized to determine the phase diagram of the $(Cu_{1-x}Zn_x)_2OSeO_3$ sample. Measurements of the electric polarization along the [001] crystal axis were measured as a function of the applied magnetic field along the [110] axis. Data measured after ZFC are shown in Fig. S2. At lower temperatures in Fig. S2A, the increase in $P$ around 0 mT is associated with the formation of the helical state. In Fig. S2B, the additional modulations in $P$ at 55 and 54 K are associated with the formation of the equilibrium skyrmion state.

The extent of the metastable SkL state was then determined by measurements of $P$ after FC. An example of the electric polarization measured as a function of field following this measurement procedure at 10 K is plotted in Fig. S3D. Here, the large increase in $P$ after FC (following paths 1 and 3) can be attributed to the formation of the metastable skyrmion state.

Such FC measurements were then repeated at a range of temperatures to determine the magnetic phase diagram shown in Fig. 2A of the main text. The full FC dataset is plotted in Fig. S2. In Fig. S2C, the measured pyroelectric current is plotted as a function of the applied magnetic field following the ZFC (thin line) and FC (thick line) measurement procedures. The signal anomaly marked with the yellow arrow is associated with the upper boundary of the metastable SkL state. The electric polarization measured as a function of the applied magnetic field after ZFC (thin lines) FC (thick lines) is plotted in Fig. S2D, where the extent of the low and high field boundaries of the metastable SkL state are indicated by the dark blue and yellow arrows respectively, and the transition of the helical to the conical state is marked by the light blue arrows.

### Details of the Lifetime Measurements

In the main text, a selection of the raw time-resolved AC susceptibility was plotted over the measured field range, featured in Fig. 3A and B. In Fig. S3A-O, the full dataset is plotted across the entire field range. Similarly, the full dataset for Arrhenius law measurements, featured in Fig. 3C and D of the main text, is plotted in Fig. S3P-T. In each case, the raw data is fitted by the stretched exponential function, as explained in the main text. For all of these lifetime measurements, the inclusion of the $\beta$ stretching parameter is necessary to successfully fit the observed decay behavior.

**Simulation Energy Minimization**

For the energy minimization of the initial states, which we call relaxation, we use the Landau-Lifshitz-Gilbert (LLG) equation. Numerically, it is necessary to fix the length of the magnetization, which changes due to the effect of error propagation during the integration. Therefore, we add a correction term to the LLG equation to keep the spin length |**s**| at unit length. According to this, our discrete spin code is implemented with the following LLG equation,

$$\frac{\partial \mathbf{s}}{\partial t} = -\gamma \mathbf{s} \times \mathbf{H}_{\text{eff}} + \frac{\alpha_G \gamma}{\mu} \mathbf{s} \times \mathbf{s} \times \mathbf{H}_{\text{eff}} + c \sqrt{\left(\frac{\partial \mathbf{s}}{\partial t}\right)^2} (1 - \mathbf{s}^2) \mathbf{s} \tag{S1}$$

where $\mathbf{H}_{\text{eff}}$ is the effective field, $\gamma$ is the gyromagnetic ratio, which sets the time scale of the integration, and $c$ is a weight for the corrector term. Since we were not interested in the dynamics of the system, we set $\gamma = 1$ and accelerated the minimization by removing the precessional term (first term to the right hand side) of Equation (S1) and by setting an appropriate damping value.

To stop the minimization process it is necessary to set a stopping criteria for the integration of Equation (S1). Thus at every time step $t$ we estimate the maximum value of the time derivative among all spins **s** using the time of the previous time step $t_{\text{prev}}$, *i.e.* we calculate

$$\Delta s = \max \left| \frac{\mathbf{s}(t) - \mathbf{s}(t_{\text{prev}})}{t - t_{\text{prev}}} \right| \tag{S2}$$

If $\Delta s$ is less than a specific threshold value $\Delta s_{\text{th}}$, we stop the minimization process. In the main study, we specify a strict tolerance by setting threshold values between $\Delta s_{\text{th}} = 10^{-5}$ and $\Delta s_{\text{th}} = 10^{-6}$, and using a large damping of magnitude $\alpha_G = 0.9$. Larger threshold values, *i.e.* a weaker tolerance, can cause the relaxation to stop at different unstable or metastable states, and does not fully guarantee that the configuration being stabilized is sitting exactly at a local minimum: the algorithm might jump to a lower energy local minimum after a sufficient number of iteration steps. On the other hand, the GNEBM can be used as an extra proof that this configuration is indeed a minimum by not showing a lower energy local minimum immediately next to the weakly relaxed state. Unless explicitly stated, the configurations used in our simulations are relaxed using a small threshold value of $\Delta s_{\text{th}} = 10^{-6}$, which means a strong tolerance within our criteria. This improves the likeness that the energy minimized state is sitting at a true local energy minimum.

**Initialization of the Simulated States**

In order to stabilize spiral states and metastable skyrmion tubes in the simulations, suitable magnetic configurations were specified as initial states and then relaxed using an over-damped Landau-Lifshitz equation. The choice of initial states is non-trivial because multiple metastable states coexist in the system model, where many of them are energetically comparable. For instance, chiral bobbers of varying length can be stabilized at sufficiently weak applied fields and characterizing all these configurations by their dimensions, location and energy becomes a challenging task. Furthermore, mapping all the possible transitions that can mediate the destruction of a skyrmion tube via the creation of Bloch points, and thus the creation of chiral bobbers, would be a time consuming project. One possible methodology would be to systematically use a parameterized mathematical function to describe chiral bobbers of different size and location and then minimize their energy, but this goes beyond the scope of this work.

The approach chosen in this study is to start from simple initial magnetic configurations. Application of the GNEBM algorithm will then find an optimal transition between, for instance, a skyrmion tube and a one dimensional modulation such as a conical phase or a spin polarized state. Of course, this does not guarantee the result of the algorithm is the only transition, but a comparison with the experimental data show excellent agreement with the obtained numerical results.

**Helical State Initialization**

One dimensional modulations are specified using spiral solutions. In the case of a helical domain propagating in the $\hat{\mathbf{k}}$-direction, which we set in the $xy$-plane forming an angle of $\phi_{\text{rot}}$ with respect to $\hat{\mathbf{z}}$, is specified by,

$$\hat{\mathbf{k}} = (\sin\phi_{\text{rot}}, 0, \cos\phi_{\text{rot}}), \tag{S3}$$

$$\hat{\mathbf{k}}_1 = (\cos\phi_{\text{rot}}, 0, -\sin\phi_{\text{rot}}), \tag{S4}$$

$$\hat{\mathbf{k}}_2 = \hat{\mathbf{k}} \times \hat{\mathbf{k}}_1. \tag{S5}$$

where the $\hat{\mathbf{k}}_i$ vectors determine the plane perpendicular to the propagation vector where the spins rotate. According to this, the spin orientation $\mathbf{s}$ is given by,

$$\Theta = \frac{2\pi}{\lambda_s}\hat{\mathbf{k}} \cdot \mathbf{r}, \tag{S6}$$

$$\mathbf{s} = \hat{\mathbf{k}}_1 \cos\Theta + \hat{\mathbf{k}}_2 \sin\Theta. \tag{S7}$$

A helical domain propagating in the $z$-direction would be obtained using $\phi_{\text{rot}} = 0$, for example. In the simulations we employ a helical domain propagating in the $x$-direction (with spins rotating in the $yz$-plane) with a periodicity of 10 lattice spacings, thus we set $\lambda_s = 10$, and $\phi_{\text{rot}} = \pi/2$. Furthermore, we found it necessary to apply a cubic anisotropy of $k_c = 0.05$ to stabilize the helical state in this orientation. The result of the minimization process using these parameters is shown in Fig. S4A. In the simulations we minimize a helical domain at zero field and perform a field sweep up to $b_z = 0.18$.

To obtain a skyrmion tube embedded in a helical background, a skyrmion profile with a radius of $r_{\text{sk}} = 5$ to 6 lattice spacings was artificially added to the previously obtained pure helical configurations. We use the following linear model for the skyrmion profile with its core oriented in the negative $z$-direction,

$$k = \frac{2\pi}{r_{\text{sk}}}$$

$$\Theta = k\rho \tag{S8}$$

$$\Psi = \varphi + \pi$$

$$\mathbf{s} = (\sin\Theta\cos\Psi, \sin\Theta\sin\Psi, -\cos\Theta) \tag{S9}$$

where $\rho$ and $\varphi$ are cylindrical coordinates. This initial state was then minimized at a range of applied field values.

There are two possible metastable states that can be obtained according to the numerical tolerances used to stop the energy minimization. By setting a stronger tolerance, which means decreasing the magnitude of the stopping criteria, the skyrmion tube starts to bend within the helical domains, in particular at weak applied field values. A result of performing energy minimization with a strong tolerance is shown in Fig. S4B. A problem with this metastable state is that the GNEBM struggles to find a smooth energy path, where multiple singularities appear in the process and which are not relevant for our study. Since we are interested in the creation of Bloch points to destroy skyrmion, the most suitable embedded skyrmion state is obtained using a weak numerical tolerance, where the resulting configuration is a straight skyrmion tube, as shown in Fig. S4C.

**Conical State Initialization**

In the case of the conical state, because the applied field is specified in the $z$-direction, we use the following initial state configuration, where $\lambda_c = 10$.

$$\cos\Theta = \frac{b_z}{d^2} \tag{S10}$$

$$\Psi = \frac{2\pi}{\lambda_c} z \tag{S11}$$

$$\mathbf{s} = (\sin\Theta\cos\Psi, \sin\Theta\sin\Psi, \cos\Theta) \tag{S12}$$

To initialize a SkT within the conical state, once again a cylindrical region with a radius of 3 lattice spins, embedded within a conical state background, was specified. The minimization result is shown in Fig. S5 where we compare the cases of with and without anisotropy. Without any cubic anisotropy, the conical background in the initial state remains after the minimization at $b_z = 0.40$, as shown in Fig. S5A. Examples of minimized configurations at a range of applied fields are shown in Fig. S5B. Without any applied cubic anisotropy, the conical state is stable over a much wider field range. Thus, we chose to utilize the no-anisotropy case when considering SkT to conical decay in the main text.

When considering a cubic anisotropy of $k_c = 0.05$, as used in the helical simulations, we still used a conical background in the initial state, although it relaxes to the spin polarized state in a large range of applied field values. According to this, skyrmion tubes appear more straight with cubic anisotropy and are observed within a conical background only for fields close to $b_z = 0.28$. Result for the cubic anisotropy case are shown in Fig. S5C and D.

**Minimized State Energies**

The energies of the helical, conical, uniformly magnetized and SkT within a helical state are shown in Fig. S5E, for the case of an applied cubic anisotropy of $k_c = 0.05$. One can see that the helical state is the lowest energy state for low fields up to $b_z = 0.10$. Between $b_z =$ 0.20 and 0.28, the conical state is the lowest energy state until the uniformly magnetized state stabilize at $b_z = 0.30$. However, over a short range of field between $b_z = 0.10$ and 0.20, the SkT within the helical state is the state with the least energy of the system. This is likely due to the small simulated system size, stabilizing the SkT state against the competing helical and conical states. The results for $k_c = 0.00$ are displayed in Fig. S5F, where below $b_z = 0.44$, the conical and SkT within the conical state have comparable energies, and the uniformly magnetized configuration only becomes the ground state above $b_z = 0.44$.

**Geodesic Nudged Elastic Band Method**

To find minimum energy paths for the different equilibrium states we use the Geodesic Nudged Elastic Band Method (GNEBM) (*13*) implemented in the Fidimag code (*48*). The algorithm minimizes the energy of a series of images along a transition path between the initial and final states by following gradients of energy in configuration space (given by the orientation of each spin site). This requires a suitable initial state for the images. Equal separation of the images is ensured by the application of a spring force. The equilibrium states are not modified. In the optimal case, the method finds a transition path that passes through one or more first-order saddle points which are a maximum in a single dimension of configuration space. The highest energy first-order saddle point between two minima of energy determines the energy barrier between the equilibrium states. To be certain that images sit at the saddle points, a climbing image GNEBM scheme was applied to the images after an appropriate number of energy minimization iterations.

Since the decay of skyrmion tubes into a one dimensional modulation involves the creation of one or multiple Bloch points, the energy landscape is rough, and multiple local energy minima exist. For example, unstable configurations such as torons usually appear mediating the transition after a skyrmion tube breaks. In our study, we have observed that the GNEBM algorithm is sensitive to the chosen dynamical equation to minimize the energy of every image in the MEP. In our code, we use a variable step size minimization that increases the numerical step size of the dynamical equation when the minimization process starts to converge or, alternatively, the energy of the images does not change significantly. On the one hand, this method causes the algorithm to converge more efficiently in most cases. On the other hand, when the energy landscape is not very smooth (for example, where the energy changes substantially close to the local minima) these large step sizes can cause the algorithm to find an alternative path by jumping over energy saddle points and thus producing an energy band with multiple peaks. In the worst cases, the algorithm does not converge properly or the resulting energy band does not have a clear first order saddle point.

Although a variable step size algorithm shows convergence issues, we have also tried implementing a fixed step size algorithm such as the Verlet algorithm implemented in the SPIRIT code (*13*, *28*), and the behavior of the minimization is slightly better but still with convergence problems. Furthermore, stopping criteria for the skyrmion transition simulations using the SPIRIT code are not completely clear from the published studies. Therefore, the complexity of the skyrmion transitions within the studied model is likely playing a major role in the struggle of recovering smooth energy transitions. Alternative algorithms that might prove useful include the string method, which could be combined with geodesic distances, as implemented in the Fidimag code, but might be slower than a variable step algorithm. Furthermore, the SPIRIT code includes optimizations to the GNEBM that can improve the distribution of images around saddle points, however suitable values of numerical parameters are required to tune this behavior optimally.

In our simulations, the cases with bad behavior occur when the applied field is sufficiently weak and the skyrmion starts to exhibit modulations along the *z*-axis. When the applied field is strong enough, the algorithm more easily finds a smooth energy path. By including a cubic anisotropy, the skyrmion tube is mostly straight within in the uniformly magnetized state, and this improves the behavior of the energy minimization.

Simulations using the GNEBM are specified using a spring constant of 1 and different stopping criteria depending on the case being analyzed. The stopping magnitude used in most of the simulations range from $10^{-5}$ down to $10^{-7}$. The initial energy path to be minimized is obtained by interpolating images between two equilibrium states, for instance, between a skyrmion tube and the conical phase. The interpolation is obtained by applying Rodrigues rotation formula (*13*) to the angles formed between corresponding spins of the two equilibrium states. It must be noticed that choosing a different initial set of images may influence the final energy path after the energy minimization with the GNEBM. Monte Carlo studies of skyrmion transitions have confirmed that the most likely disruption of skyrmion tubes is mediated by the creation of Bloch points (*8*), however mapping all the possible transitions mediated by Bloch points is a major numerical and theoretical problem. In this context, the GNEBM provides a reasonable method to tackle this challenge, and the results obtained here agree with our experimental observations and published theoretical results.

**Details of the GNEBM Simulations**

The full GNEBM dataset for the SkT to helical, SkT to conical, and SkT to uniformly magnetized state annihilation simulations are shown in Fig. S6, S7, and S8 respectively. Insets show visualizations of the magnetic configurations in select images along the MEP, in order to display the specific Bloch point decay mechanism.

In some cases, our initial attempts at determining the energy barrier for the Bloch point nucleation produced anomalous results. Specifically, in the case of the SkT to helical state annihilation at $b_z = 0.06$ in Fig. S6B, the initial energy barrier was anomalously high. We deduced that this was likely due to the limited number of images over the Bloch point nucleation, as highlighted by the red dashed box. In the conical simulations, we found that at $b_z = 0.38$, the SkT decayed by forming only one Bloch point pair, rather than by two as shown in simulations at comparable field values, as highlighted by the red box in Fig. S7B. This resulted in an anomalously high energy barrier. Similarly, we found an incorrect MEP determination for the conical simulation at $b_z = 0.44$, as shown in Fig. S7E. To rectify these simulation errors, we performed further GNEBM simulations over a shorter range of the MEP. The results are shown in Fig. S9, and the energy barriers determined in these more detailed simulations were utilized in Fig. 6 of the main text.

Regarding the transition path originally found for a field of $b_z = 0.48$: the spin rotation method to generate the initial state the algorithm found a path in which the two Bloch points stabilize in vertical alignment, generating two aligned bobbers (see Fig. S9D). With a refinement of the algorithm by minimizing the transition in over a shorter section of the MEP and interpolating new images, the algorithm found a lower energy path in which the Bloch points appear slightly off axis, thus producing two misaligned bobbers (Fig. S9E). These two saddle points are likely close in the energy landscape. For the other field magnitudes the algorithm successfully finds the lower energy transition, preferring the path where the Bloch points are misaligned.

In Fig. 5A of the main text there is a local energy minimum where a toron emerges from the creation of two Bloch points. We confirmed the local minimum character of this configuration by energy minimizing this state using the LLG equation, although using a weak tolerance. In this case there is no anisotropy that might contribute to its stability (*51*) and no edge modulations in the form of boundary conditions as in (*29*). Therefore it is possible that within the framework of our general model both the interaction of the toron Bloch points with the bobber Bloch points, and confinement effects, allow the toron to be a locally stable state.

**Surface and Internal Bloch point Nucleation**

We also decided to compare the energy barrier required to form Bloch points at the surface, and within the interior of the system. The results for simulating the energy barrier when nucleating a Bloch point at the surface of the system are displayed in Fig. S10. Insets show visualizations of the magnetic configurations at different points along the MEP.

We then compared the resulting energy barriers with the case of Bloch points being nucleated within the interior of the system, as shown in Fig. S10D. In our primary simulations, the SkT within the conical state always annihilated by forming Bloch point pairs within the interior of the system. However, interestingly, the linear fits to the data suggest that at higher fields, it may be energetically favorable for the SkT to collapse via surface Bloch point nucleation. The trend can likely be explained by considering the stability of chiral bobbers states at lower field: at higher fields, the results in Fig. S12 suggest that they are no longer stable. Thus, at higher fields, one could perhaps expect formation of Bloch points at the surface to be energetically favorable.

However, for comparison to our experimental results, the majority of Bloch point formation will be within the bulk of the sample due to the limited surface area of a single crystal. We can speculate that such surface Bloch point nucleation will be more prominent in thin lamellae or thin films. Development of new real-space imaging techniques will be essential for distinguishing between these two annihilation mechanisms in real experimental systems.

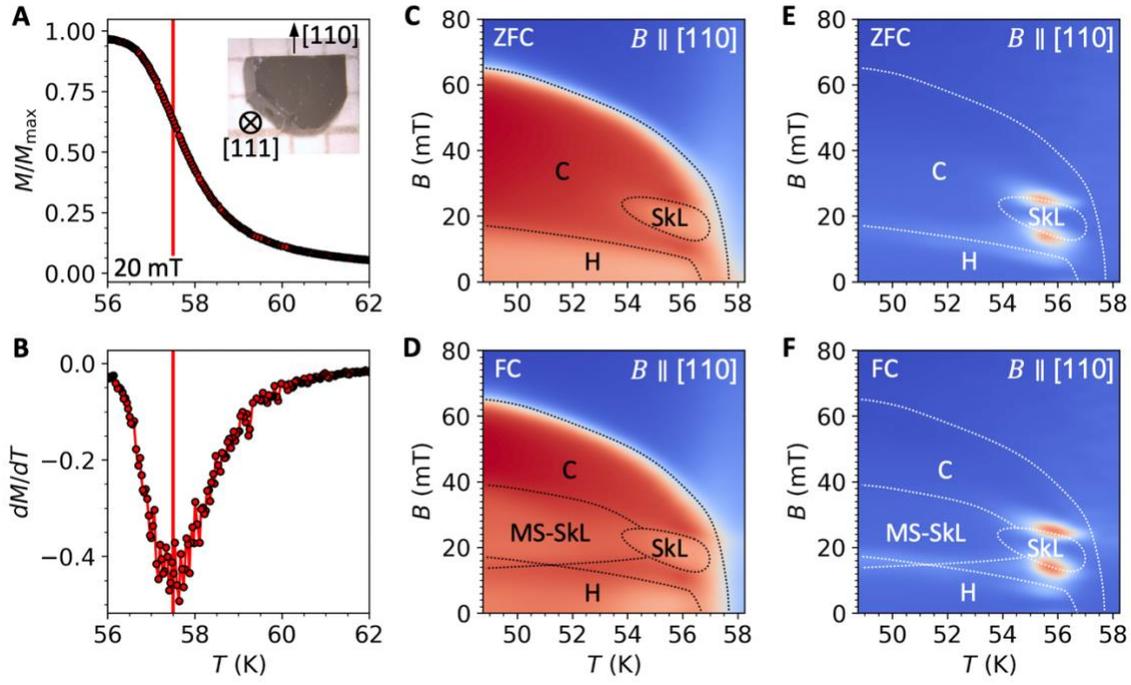

**Fig. S1: Magnetometry characterization of the $(Cu_{1-x}Zn_x)_2OSeO_3$.** (**A**) magnetization plotted as a function of temperature, measured under an applied field of 20 mT. The inset shows a photograph of the plate-shaped sample. (**B**) Numerically differentiated magnetization ($dM/dT$) plotted as a function of temperature. The vertical red line indicates $T_C$ = 57.5 K. (**C-F**) Magnetic phase diagram measured after ZFC (**C,E**) and FC (**D,F**) The colourmaps plot the real (**C,D**) and imaginary (**E,F**) components of the AC susceptibility as a function of temperature and field. Boundaries for the helical (H), conical (C), skyrmion lattice (SkL) and metastable skyrmion lattice (MS-SkL) states are indicated.

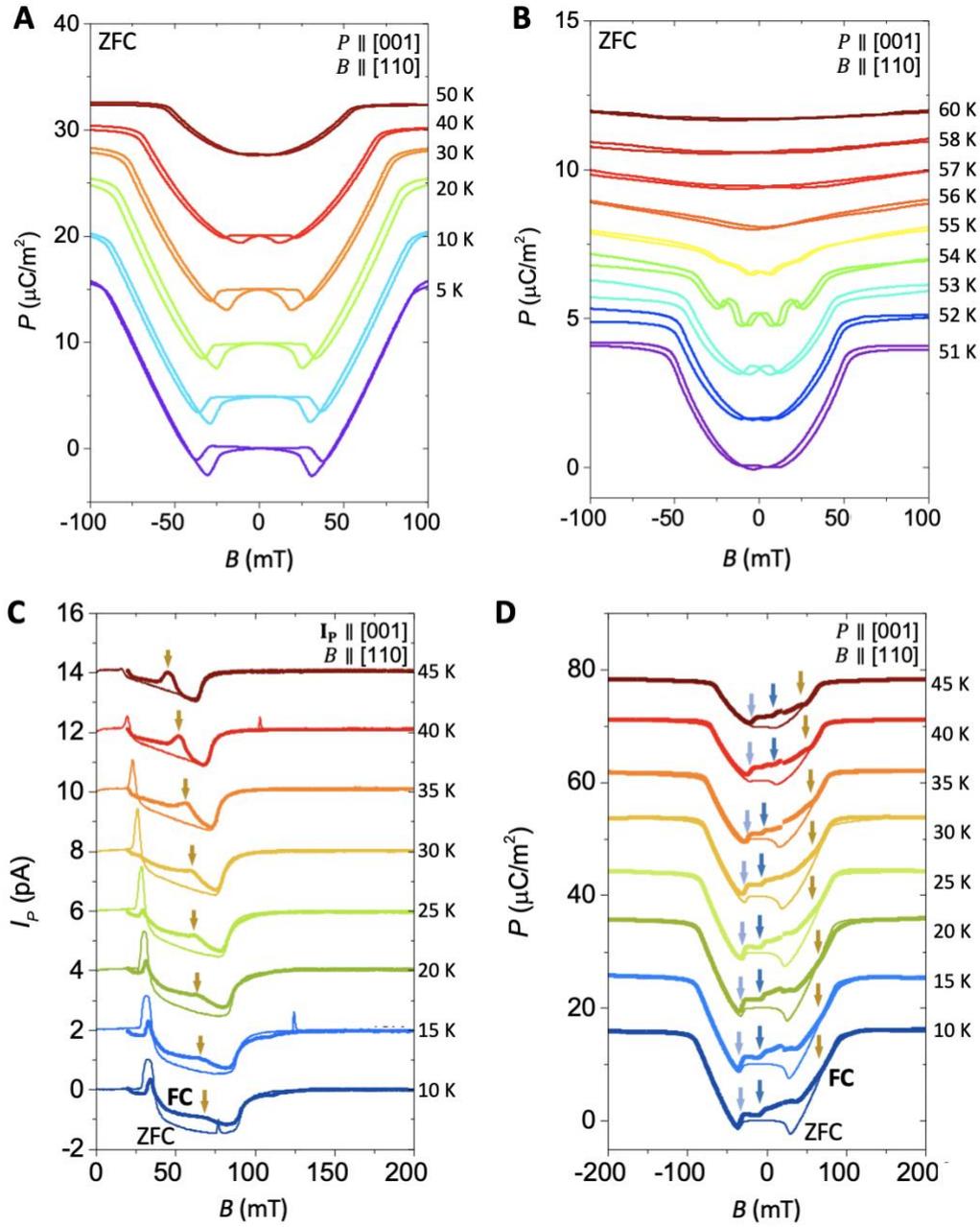

**Fig. S2: Electric polarization measurements.** (**A**,**B**) Electric polarization of the $(Cu_{1-x}Zn_x)_2OSeO_3$ sample measured along the [001] crystal axis as a function of magnetic field $B$ applied along the [110] axis at a range of temperatures. (**C**), Measurements of the pyroelectric current as a function of the applied magnetic field after ZFC (thin lines) and FC (thick lines) at different temperatures. (**D**), Measurements of the electric polarization as a function of the applied magnetic field after ZFC (thin lines) and FC (thick lines) at different temperatures. Arrows indicate the extent of the metastable skyrmion phase (dark blue and yellow), and the helical state (light blue).

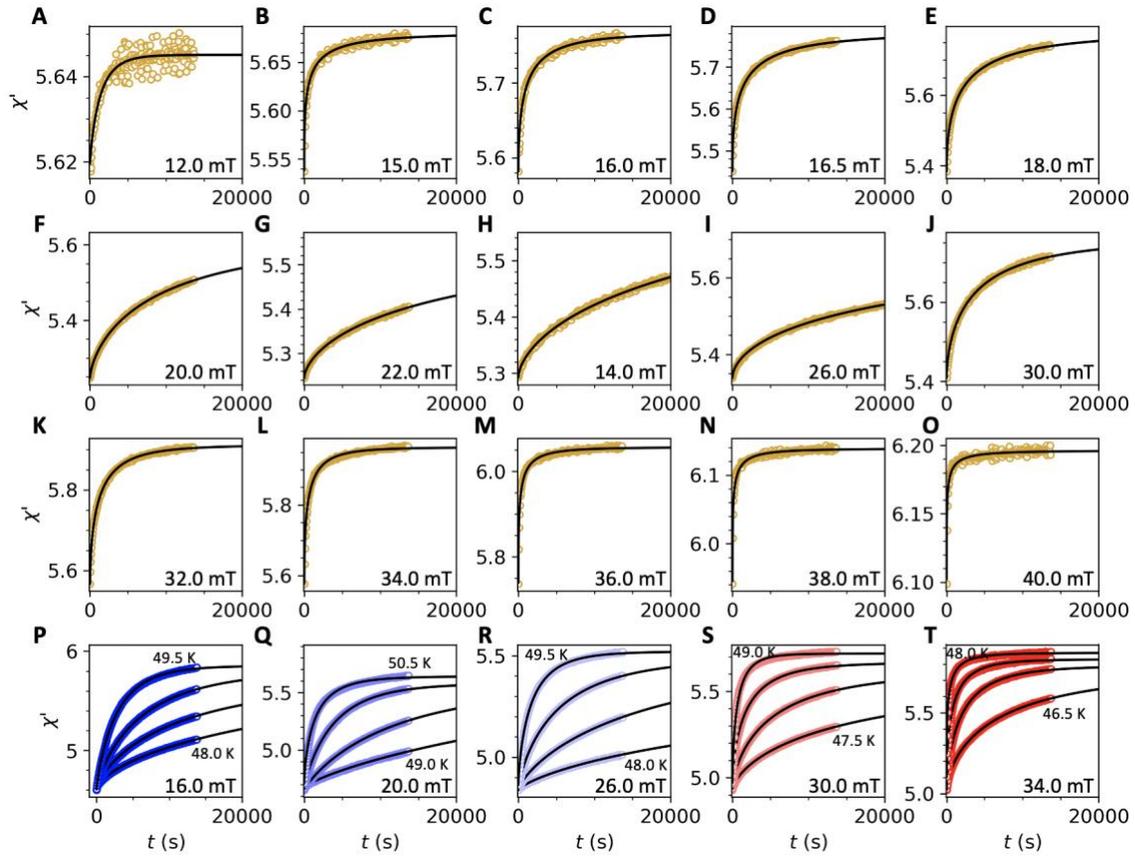

**Fig. S3: Field dependent lifetime measurements.** (**A-O**) Time resolved measurements of the real component of the AC susceptibility at range of applied magnetic fields at 49 K, performed after FC the sample through the equilibrium skyrmion phase. (**P-Q**) Time resolved measurements of the real component of the AC susceptibility at a range of temperatures and applied magnetic fields for the Arrhenius law analysis. All solid lines show fits to the stretched exponential decay model.

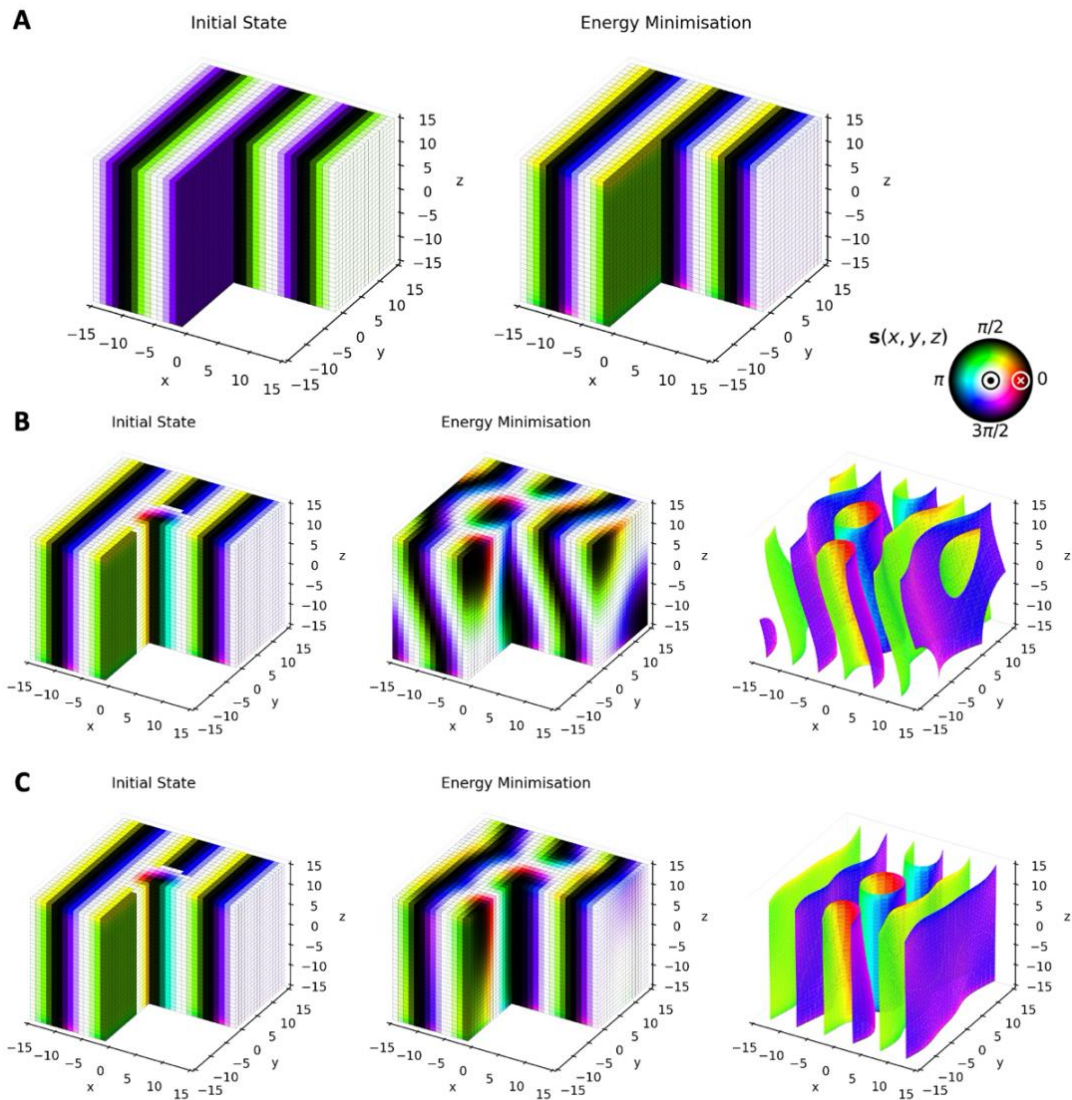

**Fig. S4: Energy minimization of the helical states.** (**A**) Initialized and energy minimized configurations of the helical domain state at $b_z = 0.00$. (**B**) Initial and energy minimized configurations of the SkT within the helical state minimized at $b_z = 0.04$ with strong convergence tolerance. (**C**) Initial and energy minimized configurations of the SkT within the helical state minimized at $b_z = 0.04$ with weak convergence tolerance.

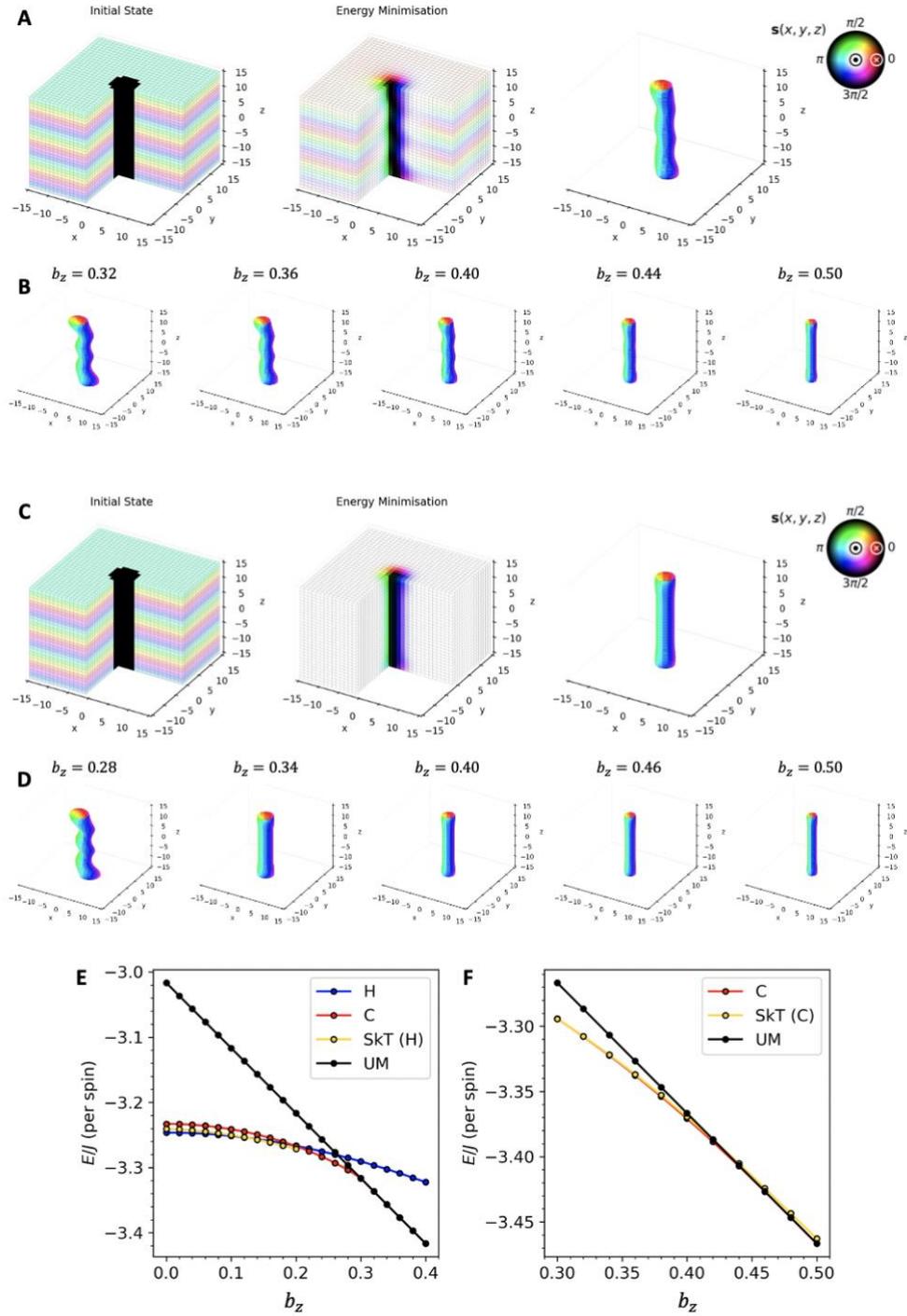

**Fig. S5: Energy minimization of the SkT within the conical and uniformly magnetized states.** (**A**) Initial and energy minimized configurations of the SkT within the conical state at $b_z = 0.40$, with no cubic anisotropy. (**B**), Minimized states at a range of applied magnetic field. (**C**) Initial and energy minimized configurations of the SkT within the uniformly magnetized state at $b_z = 0.40$, with cubic anisotropy $k_c = 0.05$. (**D**) Minimized states at a range of applied magnetic field. (**E**) Energy of the minimized helical (H), conical (C), uniformly magnetized (UM), and SkT state within the helical state (SkT (H)), plotted as a function of applied magnetic field with anisotropy $k_c = 0.05$. (**F**) Energy of the minimized conical (C), uniformly magnetized (UM), and SkT state within the conical state (SkT (C)), plotted as a function of applied magnetic field with anisotropy $k_c = 0.00$.

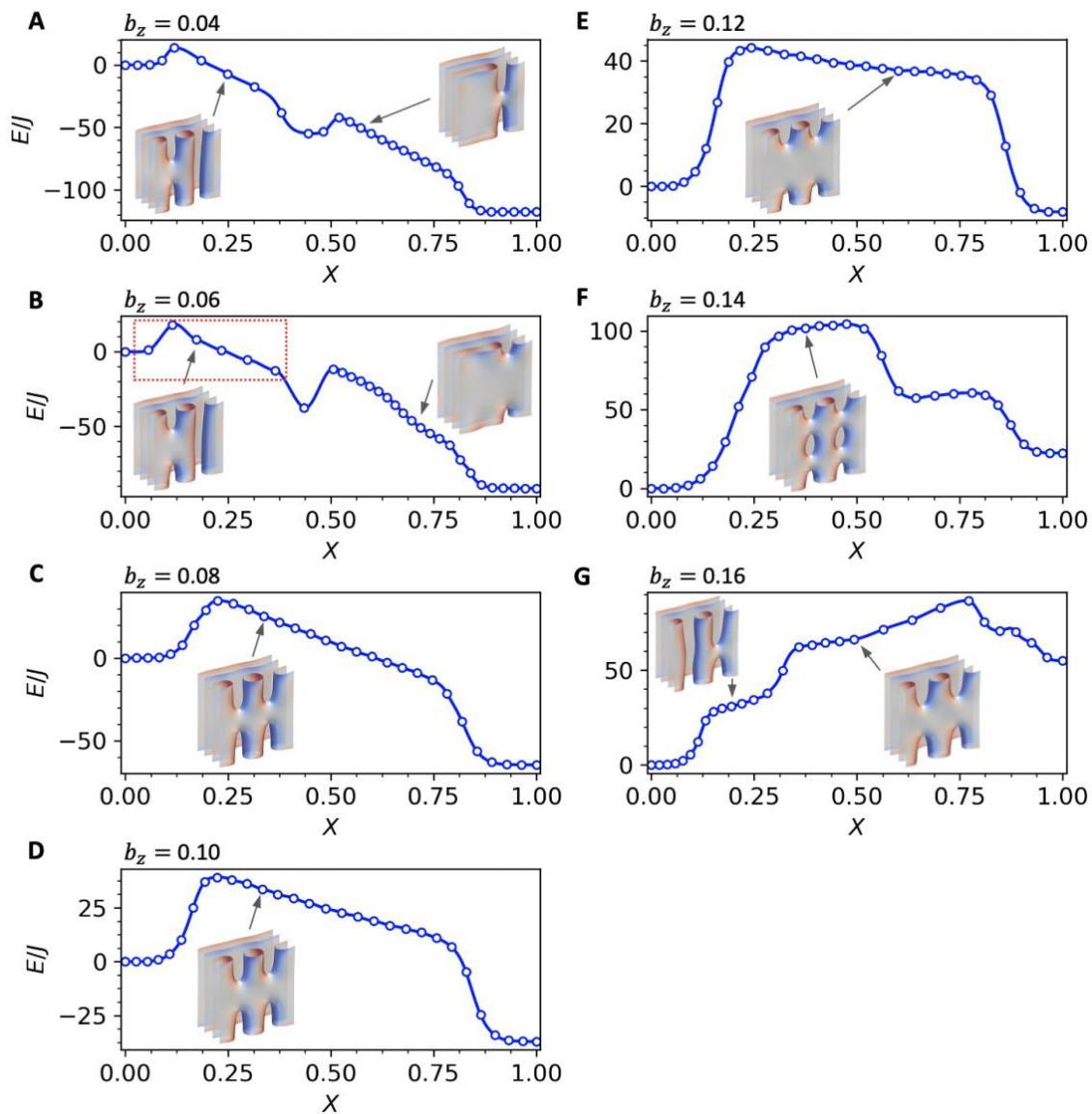

**Fig. S6: GNEBM Energy paths of the SkT annihilating within the helical state.** (**A-G**) Energy along the MEP (reaction coordinate *X*) for the SkT annihilating within the helical state at a range of applied magnetic fields with $k_c = 0.05$. The insets show visualizations of the annihilation mechanism specific to each simulation. The red dashed box indicates regions of the MEP that were repeated in more detail in the correction simulations.

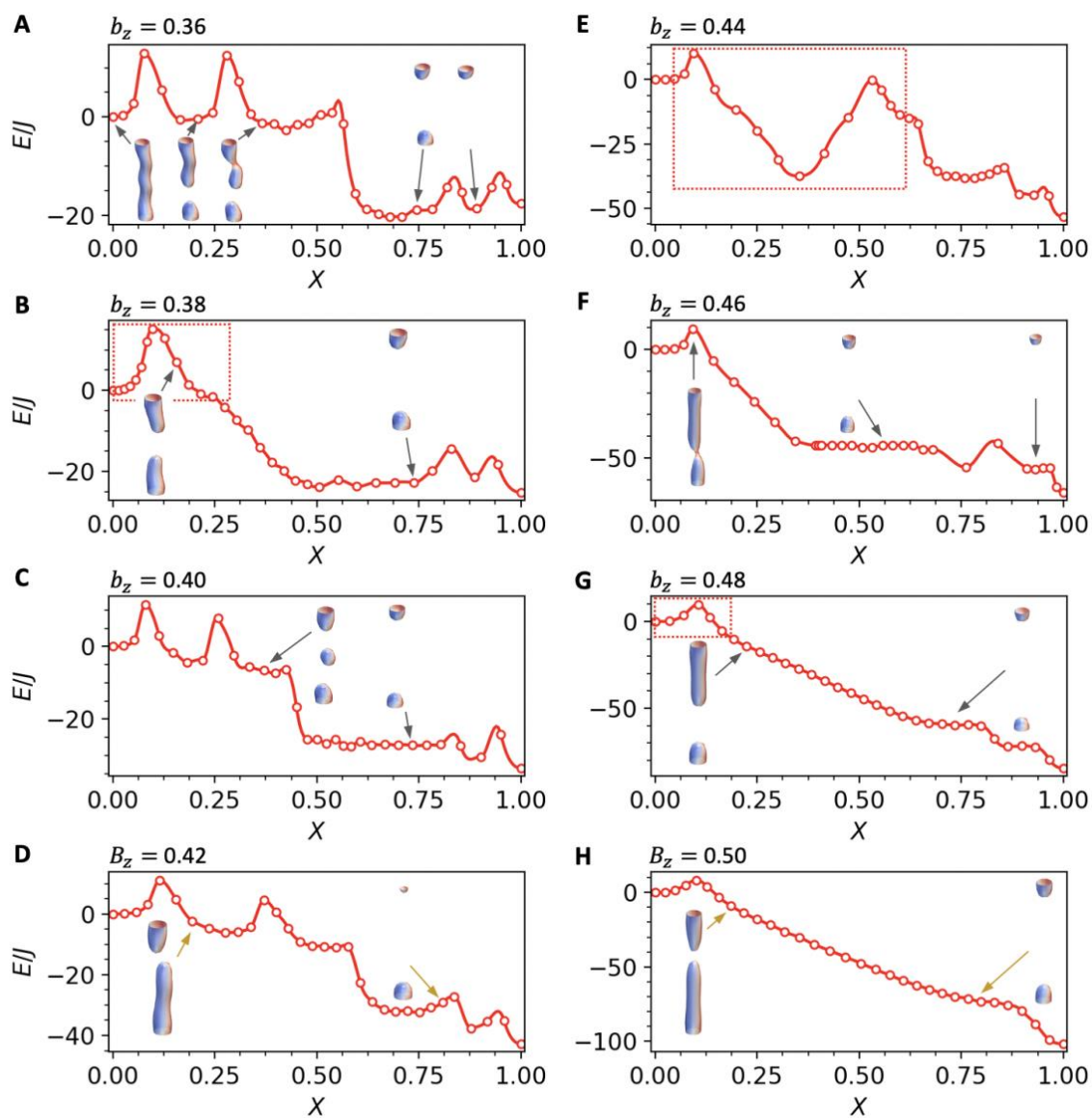

**Fig. S7: GNEBM Energy paths of the SkT annihilating within the conical state.** (**A-H**) Energy along the MEP (reaction coordinate *X*) for the SkT annihilating within the conical state at a range of applied magnetic fields with $k_c = 0.00$. The insets show visualizations of the annihilation mechanism specific to each simulation. The red dashed boxes indicate regions of the MEP that were repeated in more detail in the correction simulations.

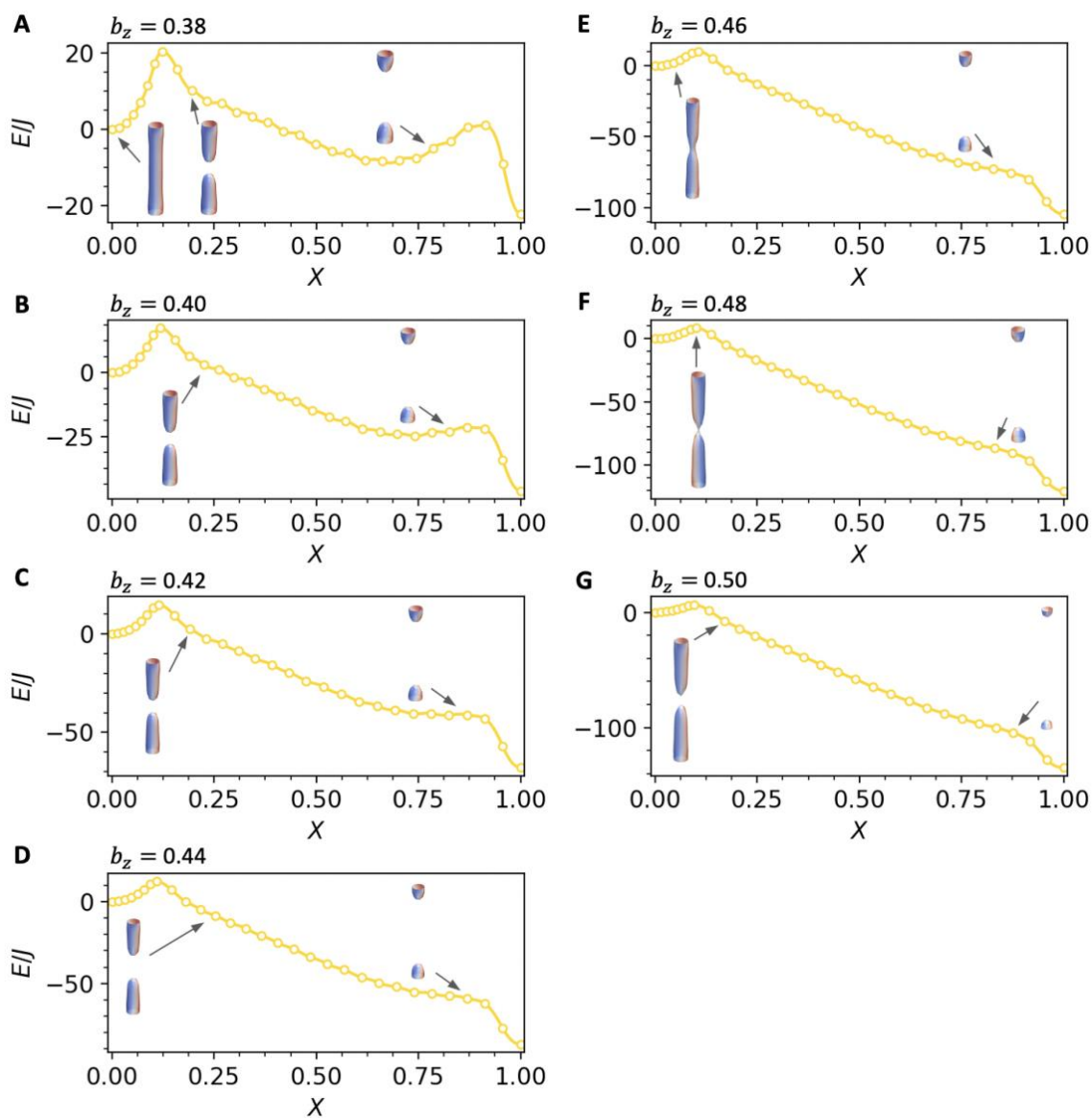

**Fig. S8: GNEBM Energy paths of the SkT annihilating within the uniformly magnetized state.** (**A**-**G**) Energy along the MEP (reaction coordinate *X*) for the SkT annihilating within the uniformly magnetized state at a range of applied magnetic fields with $k_c = 0.05$. The insets show visualizations of the annihilation mechanism specific to each simulation.

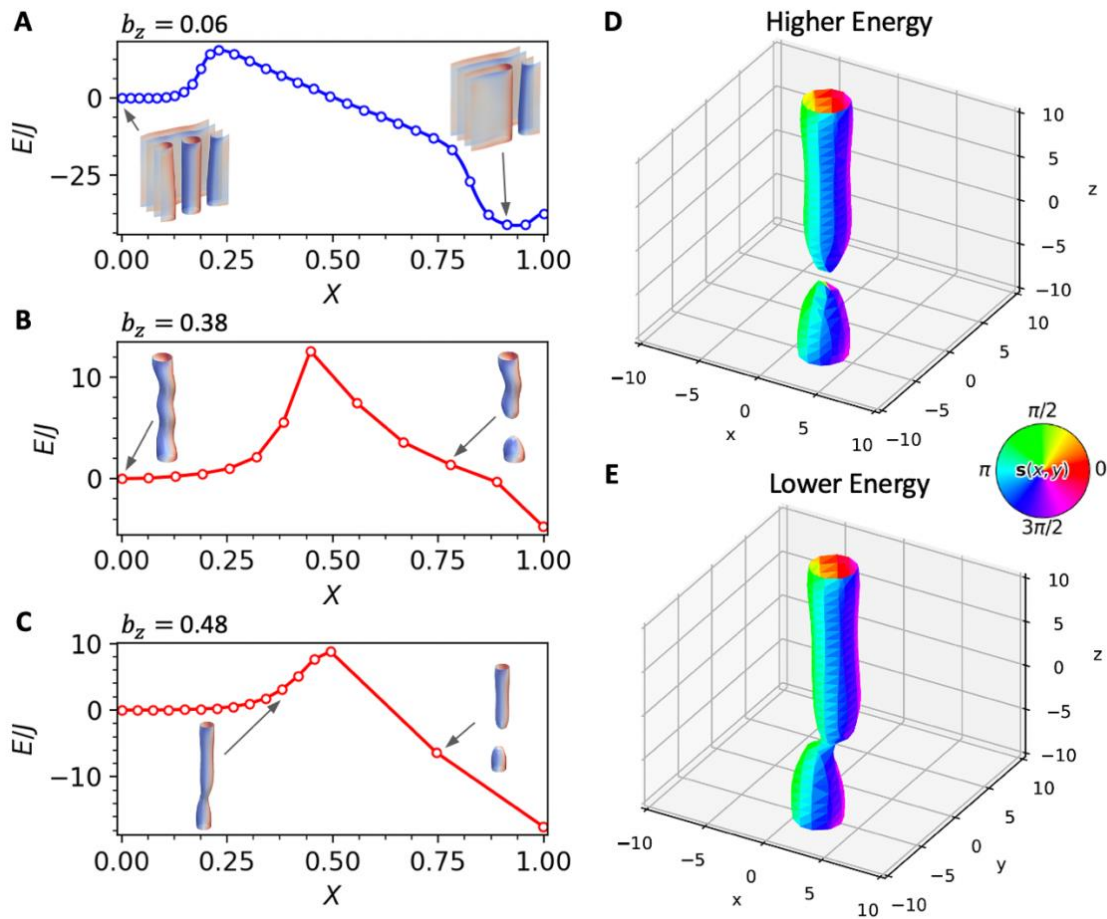

**Fig. S9: Detailed GNEBM corrections.** (**A-C**) Energy along the MEP (reaction coordinate *X*) for the SkT annihilation mechanisms within the helical state (**A**) and the conical state (**B,C**) The insets show visualizations of the annihilation mechanism specific to each simulation. (**D,E**) Two different saddle points for the annihilation of a SkT within the conical state at a field of $b_z = 0.48$.

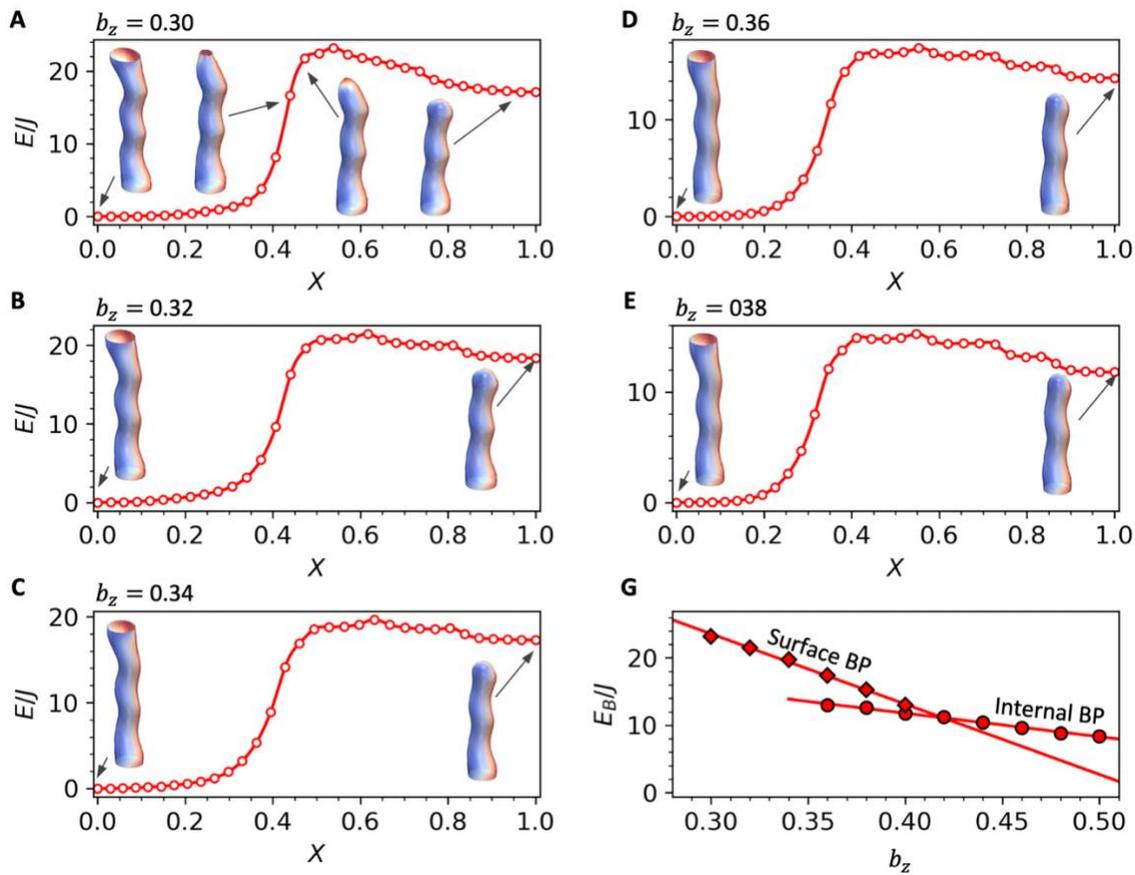

**Fig. S10: GNEBM Energy paths of the SkT annihilating within the conical state via the surface.** (**A**-**F**) Energy along the MEP (reaction coordinate *X*) for the SkT annihilating within the conical state at a range of applied magnetic fields via Bloch point nucleation at the surface of the system. The insets show visualizations of the annihilation mechanism specific to each simulation. (**G**), The simulated energy barrier required for Bloch point nucleation at the surface and within the interior of the system as a function of the applied magnetic field.